\documentclass{elsart}
\usepackage{amsmath,amssymb,curves}
\usepackage{epsfig,dcolumn}
\usepackage{graphicx}
\journal{Physics Letters B}

\newcommand{\nc}{\newcommand}
\nc{\del}{\partial}
\nc{\run}[1]{\tilde{\alpha}_{#1}}

\nc{\be}{\begin{equation}}
\nc{\ee}{\end{equation}}
\nc{\bea}{\begin{eqnarray}}
\nc{\eea}{\end{eqnarray}}
\nc{\beast}{\begin{eqnarray*}}
\nc{\eeast}{\end{eqnarray*}}
\def\xBJ{{x_{BJ}}}

\def\lsim{\mathrel{\rlap{\lower4pt\hbox{\hskip1pt$\sim$}}
    \raise1pt\hbox{$<$}}}
\def\gsim{\mathrel{\rlap{\lower4pt\hbox{\hskip1pt$\sim$}}
    \raise1pt\hbox{$>$}}}

\begin{document}
\begin{frontmatter}

\title{Exclusive electroproduction and the quark structure of the nucleon}

\author{Adam P. Szczepaniak\thanksref{DOE}}
\author{J.T. Londergan\thanksref{NSF} }

\address{ Department of Physics and Nuclear Theory Center \\
 Indiana University, Bloomington, IN 47405, USA} 

\thanks[DOE]{Supported by DOE contract DE-FG0287ER40365}
\thanks[NSF]{Supported by NSF  contract PHY0244822}

\begin{abstract}
The natural interpretation of deep inelastic scattering is in terms of hard 
scattering on QCD constituents of the target. We examine the relation between 
amplitudes measured in exclusive lepto-production and the quark content of the 
nucleon. We show that in the Bjorken limit, the natural interpretation 
of amplitudes measured in these hard exclusive processes is in terms of 
the quark content of the meson cloud and not the target itself. In this 
limit, the most efficient representation of these exclusive processes is in 
terms of leading Regge amplitudes. 
\end{abstract}


\end{frontmatter}

%%%%%%%%%%%%%%%%%%%%%%%%%%%%%%% Figure 1 %%%%%%%%%%%%%%%%%%%%%%%%%%%%%%%%%%

{\it 1. Introduction.} Recent interest in hard  exclusive lepto-production, in 
particular deeply virtual Compton scattering (DVCS) and meson 
production, has been stimulated  by the idea that these processes may give 
new insights into the quark structure of the nucleon~\cite{Ji:1998pc,Ji:1996nm,Radyushkin:1997ki,mueller-1994,mueller-1988,Goeke:2001tz,Belitsky:2001ns,Diehl:2003ny,Ji:1997gm,Burkardt:2002hr}. 
The connection between hard exclusive amplitudes and quark distributions in 
the nucleon, commonly referred to as generalized parton distributions, is 
formally analogous to that between the deep inelastic scattering cross-section 
and the structure functions. As shown by Feynman~\cite{Fey}, structure functions can be 
interpreted in terms of quark probability distributions in the nucleon. 
Duality  teaches us that, at least in principle, it is possible to use any 
channel to describe the scattering amplitude.  The parton basis of deep 
inelastic scattering (DIS) is an example of a process that is most 
efficiently interpreted in the $s$-channel representation. The basis of 
quasi-free QCD constituents is the natural choice for expressing  structure 
functions in the Bjorken scaling regime,  $Q^2 \to \infty$ and finite $x_{BJ}$. 
In this regime the 
relevant matrix elements are diagonal in the parton Fock space basis. However 
even in the case of DIS the $s$-channel parton representation becomes less 
useful in the limit $\xBJ \to 0$. In this {\it wee} parton regime it becomes 
more efficient to parametrize structure functions in terms of  amplitudes 
associated with  $t$-channel  processes. The physical  interpretation of the 
structure functions changes in between these two regimes.  As $\xBJ \to 0$ 
the structure function evolves to represent ladders of partons originating 
from $t$-channel meson exchanges.  
 
As in the case of DIS, a factorization theorem in 
exclusive lepto-production enables one to separate the hard quark-photon 
(alternatively, $W$ or $Z$ boson) scattering from the target (nucleon) 
contribution~\cite{Collins:1996fb}.  The latter contribution  is typically 
parametrized by the generalized parton distributions or GPDs~\cite{Ji:1998pc,Radyushkin:1997ki,mueller-1988}.  In analogy with deep structure functions the 
GPDs are often interpreted as corresponding to some quark distributions of 
the nucleon~\cite{Burkardt:2002hr}. Just as in the case of DIS, one can 
interpret hard exclusive lepto-production either in terms of $s$-channel 
exchanges or via $t$-channel exchanges.  

Recently Mueller and Schaefer \cite{Schafer-2006} produced a conformal spin 
expansion of GPDs.  As part of their study they investigated the extent to 
which the GPDs displayed the characteristics of their leading Regge 
trajectories.  When they examined the effective slope parameters for 
amplitudes corresponding to $\omega$ and $\rho$ exchange, they found them to 
be extremely close to the phenomenological slopes for those trajectories, a 
result they called ``quite astonishing.''  Also, a recent analysis of $\omega$ 
electroproduction with the CLAS detector at Jefferson Laboratory 
\cite{Morand:2005ex} found that their data agreed quite well with standard 
Regge phenomenology.  The purpose of this letter is to show that in the 
Bjorken limit 
%($Q^2 \to \infty$ and finite $\xBJ$), 
exclusive lepto-production amplitudes are most naturally described in terms of 
$t$-channel processes. Our results will demonstrate why one should expect 
that interpretation of the quark content of exclusive lepto-production 
processes will be most effectively discussed in the context of the parton 
content of Reggeons, rather than of the nucleon. 

Consider the case of exclusive vector meson production at high-$s$ and 
low-$t$.  As shown by a large body of evidence~\cite{Ballam:1972eq,Derrick:1996yt,Derrick:1996af,Derrick:1995vq}, such processes can be described by 
$t$-channel exchanges, where sums over exchanged mesons with all possible 
spins can be described by Regge trajectories~\cite{Irving:1977ea}.  The 
amplitude for a given Regge trajectory has the behavior $s^{\alpha (t)}$.  
At asymptotic energies, $W \gsim 10$ GeV Pomeron exchange dominates~\cite{Pichowsky:1996tn,Cano:2002zw} since it has the largest intercept $\alpha_P(0) \sim 1$
and the process is predominantly $s$-channel helicity conserving. 
% A two-gluon exchange Pomeron 
%model can also describe the high-$t$ region~\cite{Laget:1994ba,Donnachie:1988nj}. 

In this paper, we present simple arguments to justify our claim that 
hard exclusive processes are most naturally understood in terms of $t$-channel 
exchanges.  For comparison purposes, this involves a review of some very 
well known material in both deep inelastic scattering and Regge phenomenology. 
Such a review is necessary in order to compare and contrast the underlying 
mechanisms that drive inclusive and exclusive 
lepto-production in the Bjorken limit, and to clarify the differences between 
the quark/nucleon amplitudes that can be extracted from these 
reactions.  

{\it 2. The hadronic tensor in inclusive and exclusive lepto-production} 
Consider a deep inclusive reaction on a nucleon, $a^*(q) + N(p) \to  X$.  Here 
$a^*(q)$ is a virtual photon or weak gauge boson with momentum 
$q$, $N(p)$ is a nucleon with momentum $p$, and $X$ is the final state. To 
calculate the DIS cross section, one takes the square of this amplitude and 
sums over all unobserved final states $X$. As is well known~\cite{IZ}, the 
resulting inclusive cross section can be obtained from the discontinuity 
across the right hand energy cut of the forward virtual Compton amplitude.  
This is a special case of the general exclusive amplitude  
\be
  a^*(q) + N(p) \to b(q') + N(p') \ \ , 
\label{eq:exclus}
\ee
where $a^*(q)$ is a virtual boson ($\gamma, W$ or $Z$) with momentum 
$q$, where $-q^2 = Q^2$. In Eq.~(\ref{eq:exclus}), $N(p),N(p')$ represent the 
initial and final nucleons with momenta $p$ and $p'$,  respectively 
($p^2 = p'^2 = m_N^2 << Q^2$).  

Forward elastic amplitudes, which describe deeply inclusive processes, are  
characterized by $b(q') = a^*(q)$ and the kinematical relations 
$q' = q, p'=p$.  In contrast, 
for hard exclusive processes $b(q')$ is typically a real photon, meson or 
meson resonance; therefore the momentum $q'$ of $b$ satisfies $0 \le q'^2 
\sim  m_N^2 << Q^2$.  Since our main goal is to illustrate the differences 
between the hadronic contribution in exclusive and inclusive processes, in the 
following we will ignore spin and other internal degrees of freedom and 
assume only scalar currents. For these processes the hadronic contribution to 
the cross section is determined from the hadronic tensor (a scalar function 
under the above approximations), 
\begin{equation}
T = \int d^4z e^{i {{q + q'} \over 2} z} \langle p'| T \left[ j({z\over 2}) 
j(-{z\over 2})\right] |p\rangle.
\label{eq:exclusT}  
\end{equation} 
In Eq.~(\ref{eq:exclusT})  $j(z) = \phi^{\dag}(z)\phi(z)$ represents a 
(scalar) quark current in the Heisenberg picture which couples to the 
external fields 
representing the $a$ and $b$ particles. The Heisenberg nucleon states 
represent fully interacting nucleons; in particular, they include the meson 
cloud contribution.  
To study the valence and sea parton content of the nucleon the {\it bare} 
nucleon is often introduced within models that separate the QCD interactions 
among partons from chiral meson-nucleon 
interactions~\cite{Schreiber:1991qx,Holtmann:1996be}. The 
$\xBJ = {\mathcal O}(1)$  region is then found to be dominated by the bare 
nucleon and the sea contributes in the limit $\xBJ \to 0$, as expected.  
 
From Lorentz symmetry it follows that the amplitude $T$ in 
Eq.~(\ref{eq:exclusT}) is a function of four independent Lorentz scalars, 
$T=T(Q^2,\nu,t,q'^2) = T(Q^2,\xBJ,t,q'^2)$, with, $\nu = p \cdot q/m_N = 
Q^2/(2\xBJ m_N)$, and $t=(p'-p)^2 = (q - q')^2$. For inclusive processes we 
require the forward amplitude  
characterized by $t=0,0 > q'^2=q^2=-Q^2$, while the kinematics for exclusive 
processes require $t<0,0 \le q'^2 \sim m^2_N$.   Using the operator product 
expansion to leading order in QCD one finds that the matrix elements of the 
time-ordered product of the quark currents can be replaced by the product of 
two quark field operators and the quark propagator:   
%e(i(q+q')z/2) G(l) e(-ilz) <(z/2)(-z/2)> = 
%<(z/2)(-z/2)> = e(-ik_1(-z/2)) e(ik_2(z/2)) <(k_2)^* (k_1) >  G(l + (q'+q)/2) e(-ilz) <(z/2)(-z/2) > 
%delta( q/2 + q'/2 - l + k_1/2 + k_2/2)        delta( - l + k_1/2 + k_2/2)  
%l = q/2 + q'/2 + k_1/2 + k_2/2     l = k_1/2 + k_2/2  l->k
%k_2 = -/Dela/2 + k
%k_1 = Delta/2 + k
%q = -Delta/2 + m
%q' = Delta/2 + m 
\be
T(Q^2,\nu,t,q'^2) = i\int  {{ d^4 z d^4 k} \over {(2\pi)^4}} { {  e^{-ikz} } 
 \over { ( {{q + q'}\over 2} + k)^2 } } \langle  p'|T\left[ \phi^{\dag}
 ({z\over 2}) \phi(-{z\over 2})\right] |p\rangle \ .
\label{eq:T}
\ee 
Using Wick's theorem the normal ordered product of fields (in the interaction 
picture with the interaction arranged as a power series of the QCD coupling)  
was replaced by the time-ordered product since the $c$-number difference 
between the two types of ordering  does not contribute to connected matrix 
elements. The integral in Eq.~(\ref{eq:T}) is dominated by points on the light 
cone with $z^2\sim {\mathcal O}(1/Q^2)$.  It is convenient to use light cone 
coordinates, $A^{\mu} = (A^+,A^-,A_\perp)$ with $A^{\pm} \equiv A^0 \pm A^z$ 
and to choose the frame in which $q^+=0, q_\perp^2 = Q^2, p_\perp =0$.  

 For inclusive reactions where $q' = q$, the quark propagator in 
Eq.~(\ref{eq:T}) becomes  
\be
\left( {{q+q'} \over 2} + k\right)^2 = Q^2 \left( {x\over {\xBJ}} - 1 
 \right) - q_\perp\cdot k_\perp + k^2 \sim  Q^2  \left( {x\over {\xBJ}} - 1 
  \right). 
\label{eq:inclprop}
\ee 
where $x \equiv k^+/p^+$ is the fraction of the nucleon longitudinal momentum 
carried by the struck quark. The approximation is based on the observation 
that the matrix element in Eq.~(\ref{eq:T}) does not involve hard scales and 
thus on average  $k_\perp,k^- ,k^+  << |Q|$.    Under such approximations the 
absorptive part of the hadronic tensor,  $W\equiv T( \nu + i\epsilon) - 
T(\nu - i\epsilon)$ which determines the DIS cross section  is given by 
\be 
W(Q^2,\xBJ) = {1\over {Q^2}} \int \,dx \,x \delta(x - \xBJ) F(x,Q) \ , 
\label{eq:Wabs} 
\ee
where $F(x,Q)$ is the structure function,
\be 
  F(x,Q) =  {1\over 2} p^+ \int dz^- e^{-i x P^+ z^-/2 }   \langle p | T 
   \left[ \phi^{\dag}({z\over 2})\phi(-{z\over 2})\right] | p 
  \rangle_{z^+=0,|z_\perp|< {1\over {Q^2}}} . 
\label{eq:Fstruc}
\ee

For exclusive production  with  $Q^2 >> q'^2 \ge 0$, $(p - p')^2 = t <0$, 
again using light cone coordinates, to leading order in 
${\mathcal O}(Q^2)$ the quark propagator can be approximated by 
\be 
\left( {{q+q'} \over 2} + k\right)^2 = {{Q^2}\over 2}  \left( {x\over {\xi}} 
- 1\right). 
\label{eq:Qapprox}
\ee 
In the Bjorken limit $\xi = \xBJ/(2-\xBJ)$, and the longitudinal 
component of quark momentum in this case is $x = k^+/P^+$ with $P^+\equiv 
(p^+ + p'^+)/2$. 
% and $\xi$ is also given by $\xi = (p^+ - p'^+)/2$. 
The hadronic tensor for exclusive processes in the Bjorken limit is therefore 
given by, 
\be 
T(Q^2,\nu,t,q'^2) = {{P^+} \over {Q^2}}  \int  {{ dz^- d x}  \over {(2\pi)}} 
{{ i\xi e^{-iP^+ x z^-}  }   \over { x - \xi + i \epsilon} }   
 \langle  p'| T\left[ \phi^{\dag}({z\over 2}) \phi(-{z\over 2})\right] 
 |p\rangle_{z^+=0,|z_\perp|< {1\over {Q^2}}} . 
\label{eq:TDVCS} 
\ee 
The positive energy cut contribution to the hadronic tensor, which determines 
the inclusive cross section, forces $x=\xBJ > 0$. This is not the case in 
exclusive processes; here the full amplitude $T$ is needed to determine the 
cross section, so it contains an integral over both positive and negative $x$. 
Defining the free quark and anti-quark creation and annihilation operators in 
the standard way in terms of field operators, it is possible to reinterpret 
the integration over the negative-$x$ region of the quark (anti-quark) 
operator matrix elements in terms of the positive-$x$ region of the 
anti-quark (quark) operator matrix elements~\cite{Diehl:2003ny}. Thus in the 
quark representation the matrix element in Eq.~(\ref{eq:TDVCS}) receives 
contributions from pair creation and pair annihilation operators which mix 
different Fock space sectors of the nucleon wave function. 
Thus unlike DIS the DVCS matrix elements require  nondiagonal overlaps of light front wave 
functions~\cite{Brodsky:2000xy,Gardestig:2003jw}. 
%This is not the 
%case in deep inclusive scattering, except in the limit $x\to 0$. 
More detailed analysis of the correspondence between current matrix elements 
and the light cone wave function representation is given 
 in~\cite{Brodsky:2000xy}. 
% where the parton density diverges. 
We also note that calculations of 
exclusive cross sections based on GPD models that employ 
the quark handbag phenomenology also include explicit contributions from 
meson exchanges, most notably an elementary $t$-channel pion 
exchange~\cite{Mankiewicz:1998kg,Penttinen:1999th}.  

When an observable becomes sensitive to mixing between elements of a 
particular basis, it makes it difficult to interpret the 
internal structure of a state.  This suggests that for hard exclusive 
processes there may be a more efficient representation of the matrix 
elements defining the observable. In the following we will show that just 
as a hierarchy of $t$-channel processes naturally explains the low-$x$  
behavior of DIS structure functions, the same is true for the amplitudes 
 representing exclusive reactions in the {\it entire} kinematical region of 
Bjorken-$\xBJ$. 

{\it 3. $t$-channel dominance of exclusive lepto-production}. Duality implies 
that all Feynman diagrams contributing to the hadronic tensor can be 
classified as either $s$-channel exchanges with baryon quantum numbers, or 
$t$-channel exchanges with meson quantum numbers. 
For large $s$ and small $t$, $t$-channel exchange of a particle with spin 
$J$ is proportional to $\beta s^J$ with the residue $\beta$ depending on $t$ 
and particle masses (which in our case includes the large virtual photon mass, 
$Q^2$).   For example in the simple model of linear meson trajectories, the 
spin of a particle  is proportional to the square of its mass,  
$J(M^2) = \alpha(M^2)  = \alpha_0 + \alpha' M^2$  and the sum over all 
exchanged mesons leads to an amplitude proportional to $\beta s^{\alpha(t)}$. 
Such a description successfully reproduces the experimentally observed 
shrinkage of the forward peak with increasing energy~\cite{Irving:1977ea}. In 
general in the regime where $s/|t| >> 1$ the singularity $J=\alpha(t)$ 
furthest to the right in the complex angular momentum plane, determines the 
leading power of the energy dependence. We are concerned with reactions 
that obey the constraints $s/|t| >> 1$.  In this kinematic region one  
would expect $t$-channel exchanges to accurately  
parametrize these amplitudes. It is well known that in the case of DIS the 
Regge parametrization is relevant when $\xBJ \to 0$; however, for finite 
$\xBJ$ the $t$-channel exchange description becomes inefficient. This occurs 
because away from the forward region, all singularities in the complex angular 
momentum plane {\it i.e.} all daughter trajectories contribute equally to the 
amplitude as the rightmost singularity, which defines 
the leading Regge trajectory.  For DIS processes as one goes away from the 
region $\xBJ \sim 0$, it very quickly becomes more efficient to represent the 
amplitude by $s$-channel exchanges of quasi-free partons. However, we will 
show that the conditions that characterize exclusive production are quite 
different from the conditions governing the inclusive processes.  
  
\begin{figure}
\includegraphics[width=5in]{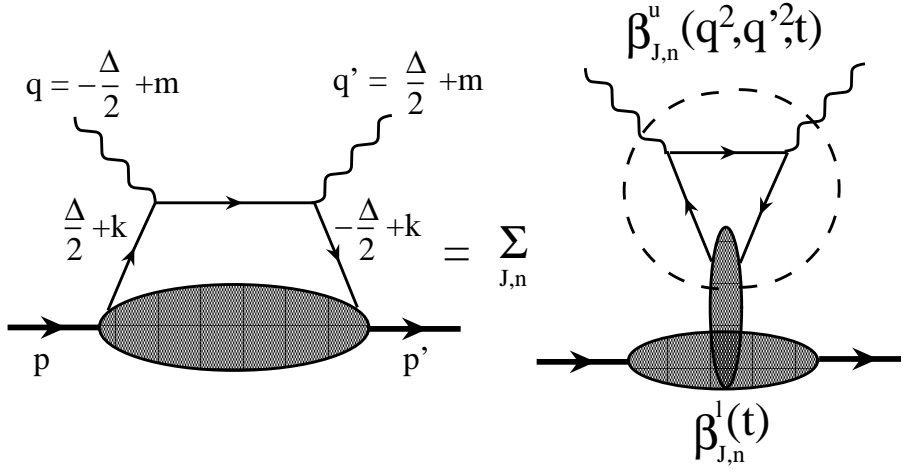}
\caption{\label{Fig:llog}  $t$-channel meson contribution to the hadronic 
tensor for exclusive lepto-production. The amplitude is summed over all spins 
$J$ that can contribute, and depends on the product of the residue functions 
$\beta$ at the upper and lower vertices.}
\end{figure}

The contribution to the hadronic tensor from $t$-channel exchange of a 
spin-$J$ meson is proportional to
\bea   
  T_J =  {{ \beta^l_{J}(t) \beta^u_{J}(q^2,q'^2,t) }\over {t - M_J^2} }   
 \sum_{\lambda=-J}^J 
&&   \left[ {{(p' + p)^{\mu_1}}\over 2} \cdots {{(p'+p)^{\mu_J}}\over 2} 
  \epsilon^\lambda_{\mu_1 \cdots \mu_J} (p'-p) \right] 
   \nonumber \\ 
 & \times &    \left[{{(q' + q)^{\nu_1}}\over 2} \cdots 
 {{(q'+q)^{\nu_J}}\over 2} \epsilon^{*\lambda}_{\nu_1 \cdots \nu_J} (p'-p) 
 \right]. \nonumber \\
\label{eq:TJ}
\eea 
 In Eq.~(\ref{eq:TJ}), $\epsilon$ is the spin-$J$ polarization vector, and 
$\beta^l_{J}$ and $\beta^u_{J}$ are the residue functions at the {\it lower} 
and {\it upper} vertex, respectively. This is shown schematically in 
Fig.~\ref{Fig:llog}. In the Bjorken limit, 
 $s \to Q^2(1-\xBJ)/\xBJ$ and the amplitude reduces to 
 \be 
  T_J =  {{ \beta^l_{J}(t) \beta^u_{J}(q^2,q'^2,t) }\over {t - M_J^2} }  
  \left( {Q^2 \over 2\xBJ } \right)^J .
\label{eq:TJred}
\ee

The key question is how the upper residue function depends on the large 
variables ($Q^2$ and $-q'^2 = Q^2$ in the case of inclusive processes, and 
$Q^2$ for the exclusive amplitudes). It is well known that for kinematics 
that are relevant to inclusive scattering, the upper residue function 
behaves as $(1/Q^2)^{J+1}$, {\it modulo} logarithmic corrections, so that the  
amplitude scales,  $ Q^2 T_J  \propto  (1/\xBJ)^{J}$, as 
expected~\cite{Ioffe,Brodsky:1971zh,Brodsky:1972vv,Brodsky:1973hm}. 
Summing over all spins leads to the Regge behavior, $Q^2 T = \sum_J T_J 
\propto (1/\xBJ)^{\alpha(0)}$.  The leading Regge trajectory with 
$\alpha(0)>0$ will dominate the $\xBJ \to 0$ behavior of the hadronic tensor, 
while all daughter trajectories with $\alpha_n(0) < \alpha(0)$ are subleading 
as $\xBJ \to 0$.  For finite $\xBJ$, however, daughter Regge trajectories are 
no longer suppressed, and as a result the Regge description becomes 
ineffective while the $s$-channel parton model description becomes natural.

We will now show that for exclusive amplitudes the upper vertex scales with a 
finite power of $1/Q^2$ instead of being suppressed for high spins. Thus after 
summing over all spins it gives an amplitude that  behaves as $T 
 \propto (Q^2/\xBJ)^{\alpha(0)}$ , {\it i.e.} so long as $Q^2 >> m^2_N$ is in 
the Bjorken regime, the amplitude is dominated by the leading Regge trajectory 
for {\it all } $\xBJ$ and not only in the limit $\xBJ \to 0$.
   
To show that, we first rewrite the contribution of a $t$-channel, spin-$J$ 
exchange to the matrix element  in Eq.~(\ref{eq:T})  in terms of the 
two-current correlation in the exchanged meson, 

\bea
& & \int d^4 z e^{-ikz}   \langle p'| T \left[ \phi^{\dag}\left( {z\over 2} 
\right)\phi\left(-{z\over 2}\right) \right] |p\rangle = 
{{\beta^u_{Jn}(t) } \over {t - M_{Jn}^2}}  \Phi_{Jn}(p - p',k)
\nonumber \\ 
& & 
 \sum_{\lambda=-J}^J  \left[ k^{\nu_1} \cdots k^{\nu_J} 
 \epsilon^\lambda_{\nu_1 \cdots \nu_J} (p'-p) \right]^*    
 \left[{{(p' + p)^{\mu_1} }\over 2}\cdots {{(p'+p)^{\mu_J}}\over 2} 
 \epsilon^\lambda_{\mu_1 \cdots \mu_J} (p'-p) \right] .\nonumber \\
\label{eq:ME} 
\eea
In Eq.~(\ref{eq:ME}), $n$ refers to other quantum numbers that distinguish 
between exchanged mesons (after Reggeization $n$ distinguishes between the 
leading and daughter trajectories), and $\Phi_{Jn}(\Delta,k)$ is the 
covariant (Bethe-Salpeter) amplitude of a meson with momentum  
$\Delta$, where $\Delta^2=t$. Finally $k$ is the relative momentum between 
the quarks, as shown in Fig.~\ref{Fig:llog}. Using dispersion relations, the 
(unnormalized) Bethe-Salpeter amplitude can be represented 
as~\cite{Nakanishi:1969ph},
\be
 \Phi_{Jn}(\Delta, k)  = i\int_{-1}^{1} dx \int d\mu^2 {{g_{Jn}(x,\mu^2) } 
 \over { \left[ (k - {x\over 2} \Delta)^2 - \mu^2 + 
 i\epsilon\right]^{n + J} } } \ ,  
\label{eq:BS} 
\ee 
 where the spectral density $g_{Jn}$ is related to the parton distribution 
amplitude in a meson and can in principle be constrained from electromagnetic 
data~\cite{Szczepaniak:1993kk} and QCD asymptotics~\cite{Lepage:1980fj}. 

What is important for our argument are the following model independent 
features of the amplitude in Eq.~(\ref{eq:BS}). The magnitude of the relative 
momentum $k$ is of the order of the hadronic scale $\mu~m_N$. Secondly, in the 
infinite momentum frame, $k \propto \Delta$, $\xi_{\pm} \equiv  (1\pm x)/2$ 
represents  the fraction 
of the longitudinal momentum carried by the quark (antiquark), and $g$ becomes 
the parton distribution function. In the Bjorken limit the leading, 
helicity-zero component of the meson distribution amplitude has 
$J$-independent behavior near $\xi_{\pm} \to 1$~\cite{Chernyak:1983ej}.  
Finally the power dependence  of the relative momentum is constrained by the 
angular momentum {\it i.e.} the power of the denominator in Eq.~(\ref{eq:BS}) 
increases with $J$.  Inserting the analytical expression for the 
Bethe-Salpeter amplitude of Eq.~(\ref{eq:BS}) into Eq.~(\ref{eq:ME}) and then 
into Eq.~(\ref{eq:T}), one obtains the final expression for the contribution 
of spin-$J$ exchange to the hadronic tensor. It is given in terms of an integral over $k$ (see Eq.~(\ref{eq:T})) of the product of the quark propagator, the 
Bethe-Salpeter amplitude of Eq.~(\ref{eq:BS}), and a polynomial in $k$ 
originating from the coupling to the spin-J polarization vectors 
(Eq.~(\ref{eq:ME})). The polynomial is responsible for the $s^J \sim (Q^2)^J$ 
behavior of the amplitude. The integral can easily be evaluated using the 
Feynman parametrization which introduces an integral over the parameter 
$\alpha$. Ignoring terms of order $m_N^2/Q^2$ and $t/Q^2$, up to constant 
numerical factors one finds    
\bea
\beta^u_{Jn}  &=&  \int_{-1}^1 dx \int d\mu^2  g_{Jn}(x,\mu^2) \int_0^1 
d\alpha {{ \alpha^J } \over {\left[ -\alpha \left(  {{q'^2 + q^2} \over 2 }  
 + x{{ q'^2 - q^2} \over 2}\right) + \mu^2\right]^{n+J-1}  } } \ .
\nonumber \\ 
\label{eq:beta} 
\eea
For inclusive amplitudes, $q'^2=q^2=-Q^2$ the $x$ disappears from the 
denominator and the integration over $\alpha$ is 
dominated by $\alpha\sim \mu^2/Q^2$.  As a result, the entire integral is of 
order $(\mu^2/Q^2)^{J+1}$, as expected. However for exclusive amplitudes, 
$q'^2 \sim 0$  the integrand is dominated by the region 
$1-x = {\mathcal O}(\mu^2/Q^2)$, and finite $\alpha$. The endpoint behavior of 
the distribution amplitudes $g_{nJ}$ is spin independent, and for 
leading-twist amplitudes  $g_{Jn}(x \to 1) \sim (1-x)$.  This leads to 
a $J$-independent suppression of the upper vertex with $Q^2$, $\beta^u_{Jn} 
\sim {\mathcal O}(\mu^4/Q^4)$ {\it which is independent of the spin of the 
exchanged meson}  . This is our main result. As discussed above, 
upon summing over all spins from a single trajectory one determines that the 
hadronic tensor is proportional to 
$(Q^2/\xBJ)^{\alpha(t)}\sim (Q^2/\xBJ)^{\alpha(0)}$, for small $t$. 
 Thus, in the  Bjorken limit hard exclusive processes should be dominated by 
a single, leading Regge trajectory for all $\xBJ$, and not just for 
$\xBJ \to 0$.  We argue that this is the most efficient way to interpret hard 
exclusive processes.  We also note that the Regge approach to exclusive deeply 
virtual production was previously considered in~\cite{Korenblit:1979cw}, where 
a different $Q^2$ dependence was obtained for the full exclusive amplitude. 
However, those authors assumed a Regge-like amplitude with a particular $Q^2$ 
dependence, rather than deriving the behavior from a sum of $t$-channel poles 
as was done here. 

As we mentioned earlier, a recent experimental analysis of $\omega$ 
electroproduction at Jefferson Laboratory~\cite{Morand:2005ex} 
showed that their data was in good agreement with predictions from standard 
Regge phenomenology, while showing large uncertainties with analyses based on 
models of generalized 
parton distributions~\cite{Diehl:2005gn}. Note that our results 
were derived in the Bjorken limit with $s/|t| >> 1$, while the JLab results 
correspond to energies of a few GeV and values up to $|t| \sim 2.7$.  
%In this region we do not expect our predictions to be valid.  
At high energies, one expects $\omega$-photoproduction to be dominated by 
Pomeron exchange; however at lower energies the leading meson Regge 
trajectories $f_2$ and $\pi$, with intercepts $\alpha_{f_2}(0) \sim 0.5$ and 
$\alpha_{\pi}(0) \sim 0$ can also give sizeable contributions.  This indeed 
was found to be the case for the CLAS data~\cite{Morand:2005ex,Laget:2004qu}. 
These exchanges are also found to be responsible for $s$-channel helicity-flip 
amplitudes.  Extension to $|t| \gsim 1\mbox{ GeV}^2$ is somewhat model 
  dependent~\cite{Donnachie:1988nj,Laget:1994ba} as one needs to extrapolate 
further from the  physical region of the $t$-channel.  In all these analyses, 
once $W$ is greater than a few GeV, the daughter Regge trajectories with 
$\alpha_n(0)<0$ do not seem to be needed.  

 {\it Summary} 
We have shown that in the Bjorken limit ($s >> |t|$ and $Q^2 >> m_N^2$), the 
leading Regge trajectory should be expected to dominate amplitudes for 
exclusive lepto-production. Our arguments also imply that the generalized 
parton distributions can be written in terms of Reggeon-nucleon coupling and 
that their `natural' interpretation would be in terms of the parton content 
of the meson cloud rather than that of the bare nucleon.  The GPDs can be 
computed by summing amplitudes of the type given in Eq.~(\ref{eq:ME}) with  
various sum rules constraining the products of residue functions and 
Reggeon-quark-antiquark distribution amplitudes~\cite{Polyakov:2002wz}.  
Our arguments are model-independent and are based on general assumptions about 
the analytic structure of the scattering amplitude in the complex angular 
momentum plane. However, analyses of hard exclusive processes, particularly 
those at relatively low energies and large $|t|$ values, will require detailed 
models that can accommodate spin-flavor dependence and build in the  
characteristics of the relevant Regge trajectories. The recent analysis of 
$\omega$-photoproduction~\cite{Morand:2005ex} indeed suggests that Regge 
phenomenology can successfully be used to describe the hadronic part of the 
production amplitude in exclusive lepto-production. 

We would like to thank Stan Brodsky, Jean-Marc Laget, Wally Melnitchouk, 
Dieter Mueller, Anatoly Radyushkin, and Christian Weiss for useful comments 
and discussions.

\end{document}